\documentclass[graybox]{svmult}
\usepackage{mathptmx}
\usepackage{bm,amsfonts}       
\usepackage{helvet}         
\usepackage{courier}        
\usepackage{type1cm}        
\usepackage{makeidx}         
\usepackage{graphicx}        
\usepackage{multicol}        
\usepackage[bottom]{footmisc}

\newcommand{\beq}{\begin{equation}}
\newcommand{\eeq}{\end{equation}}
\newcommand{\beqa}{\begin{eqnarray}}
\newcommand{\eeqa}{\end{eqnarray}}
\newcommand{\nn}{\nonumber \\}
\newcommand {\np}[1]{{\mbox{\textrm{:}\,}{#1}{\,\textrm{:}}} }

\def \b {\beta}

\def \disk {\mathrm{disk}}
\def \e {\mathrm{e}}
\def \el {\textrm{\scriptsize{el}}}

\def \la {\langle}
\def \ra {\rangle}
\def \s {\sigma}
\def \t {\tau}

\def \I {{\mathbb I}}

\def \Z {{\mathbb Z}}
\def \ch {\mathrm{ch}}

\def \z {\zeta}
\def \L {\Lambda}
\def \D {\Delta}
\def \A {\mathcal A}
\def \J {\mathcal J}

\def \Im {\mathrm{Im} \, }
\def \mod {\ \mathrm{mod} \ }
\def \H {{\mathcal H}}
\def \tr {\mathrm{tr}}
\def \uu {{\widehat{u(1)}}}

\def \W {\mathcal{W}}
\begin{document}

\title*{Hilbert space decomposition for Coulomb blockade in  
Fabry--P\'erot interferometers}
\titlerunning{Hilbert space for Coulomb blockade} 
\author{Lachezar S. Georgiev}
\authorrunning{L.S. Georgiev} 
\institute{Lachezar S. Georgiev \at Institute for Nuclear Research and Nuclear Energy, 
Bulgarian Academy of Sciences, \at
72 Tsarigradsko Chaussee, 1784 Sofia, Bulgaria, EU; 
\email{lgeorg@inrne.bas.bg}}

\maketitle

\abstract{
We show how to construct the thermodynamic grand potential of a droplet 
of incompressible fractional quantum Hall liquid, formed inside of an electronic 
Fabry--P\'erot interferometer, in terms of the conformal field theory disk partition 
function for the edge states in presence of Aharonov-Bohm flux. 
To this end we analyze in detail the algebraic structure of the edge states' Hilbert space 
and identify the effect of the variation of the flux. This allows 
us to compute, in the linear response approximation, all thermodynamic properties 
of the conductance in the regime when the Coulomb blockade is softly lifted by the 
change of the magnetic flux due to the weak coupling between the droplet and the 
two quantum point contacts. 
}
\section{The FQHE  Fabry--P\'erot interferometer}
\label{sec:1}
The electronic version \cite{Mach-Zehnder-Nature} of the famous optical  Fabry--P\'erot interferometer, 
which we will analyze here,  is constructed by two quantum point contacts (QPC) inside of an 
incompressible fractional quantum Hall (FQH) bar \cite{wen,fro-zee,ctz}.
In the weak-backscattering regime, small gate voltages on the QPCs 
create constrictions inside the incompressible FQH liquid and facilitate 
tunneling of non-Abelian quasiparticles along the QPCs.
However, this regime is unstable in the sense of the renormalization group flow, 
i.e., even a small number of quasiparticles tunneling along the QPCs at low $T$
significantly renormalizes the tunneling amplitudes thus intensifying  tunneling and 
eventually the two QPCs  pinch off, which corresponds to the strong backscattering regime
that is already a stable  fixed point of the renormalization group flow.

\begin{figure}[htb]
\centering
\includegraphics[bb=40 220 560 620,clip,width=6.5cm]{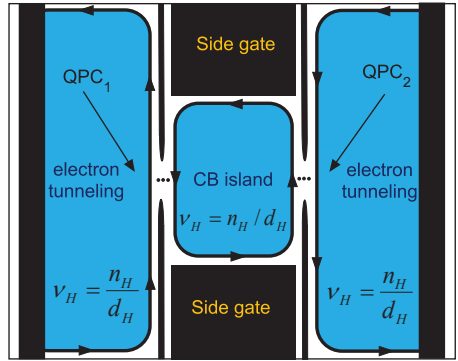}
\caption{A FQH bar with two QPCs in the strong backscattering regime in which single electrons could 
tunnel, if small bias is present,  between the three disconnected liquids producing discrete peaks in the 
conductance. The side gates' voltage could change the area of the CB island varying in this way the 
flux through the island. \label{fig:FPI-CB2}}
\end{figure}

In the strong backscattering regime, when the QPCs gate voltages are big enough 
that the two constrictions are completely pinched-off, the two-dimensional electron gas 
is split into three disconnected FQH liquids forming a 
\textit{Coulomb blockade} (CB) island in the middle, see Fig.~\ref{fig:FPI-CB2}.
Only electrons could tunnel between the disconnected parts of the interferometer
and the main mechanism at low temperature and low bias is through 
\textit{single electron tunneling}.
The conductance in the CB regime is determined in the following steps
(sequential tunneling through CB island):
first one electron  tunnels from the left FQH liquid 
through the left QPC  to the island, then the electron which is accommodated at the edge 
of the CB island is transported along the edge and then it tunnels through the right QCP  
to the right FQH liquid. Using the Landauer formula 
one can see \cite{thermal} that the CB conductance is 
\beq\label{G_CB}
G_{\mathrm{CB}} (T,\phi)=\left(\frac{h}{e^2}\right) \frac{G_L G_R}{G_L+G_R} G_{\mathrm{is}}(T,\phi)  ,
\eeq
where the CB island's conductance $G_{\mathrm{is}}$ depends on the magnetic flux $\phi=B.A$: 
for most values of the flux we have Coulomb blockade ($G=0$) and for special discrete 
values of flux we have conductance peaks \cite{stern-halperin,Stern-CB-RR,CB,nayak-doppel-CB,cappelli-viola-zemba}. 
The  tunneling conductances of the two QPCs are 
independent of the flux and vanishing at low-temperature as
$G_{L,R}\propto T^{4\D -2}$ where $\D$ is the scaling dimension of the electron operator.
\section{Coulomb blockade  island's conductance--the CFT point of view}
\label{sec:2}
An interesting observation in this setup is that the conductance of the CB island can be 
explicitly computed at finite temperature within the framework of the conformal field theory \cite{thermal}.
This is due to the Einstein's relation \cite{DiVentra,thermal}, which expresses the conductance 
$\s(0)$ in terms of the charge stiffness (or, thermodynamic density of states) 
\beq \label{Einstein}
\s(0)=e^2 D \left. \frac{\partial n}{\partial \mu} \right|_{T}, 
\eeq
where $D$ is the diffusion coefficient,   $\mu$ is the chemical potential,
$n$ is the electron density and the thermodynamic derivative  is at constant temperature.

The diffusion coefficient is usually related to the relaxation time \cite{DiVentra},
for normal conductors, however, for a \textit{ballistic one-dimensional  channel}, 
such as the FQH edge, the relaxation time 
must be replaced by the \textit{time-of-flight} $\tau_f$ and the diffusion coefficient
could be written as follows  \cite{DiVentra}
\[
D_{\scriptsize\textrm{bal}}= v_F^2 \tau_f, \quad \t_f=L/(2v_F) \ \Rightarrow \ D=Lv_F/2 ,
\]
where $v_F$ is the Fermi velocity at the edge and $L$ is the circumference or length of the edge.
According to Eq.~(\ref{Einstein}) the charge stiffness can be computed as a derivative 
of the thermodynamic average of
the particle number. To this end we shall use the Grand canonical partition function
for a disk-shaped CB island derived within the  CFT framework \cite{CFT-book,thermal}
\beq \label{Z}
Z_{\disk}(\t,\z) = \mathrm{tr}_{ \H_{\mathrm{edge}}} \ \e^{-\beta (H-\mu N)} 
= \mathrm{tr}_{ \H_{\mathrm{edge}}} \ \e^{2\pi i \t (L_0 -c/24)} e^{2\pi i \z Q}, 
\eeq
 where the Hamiltonian of the disk $H=\hbar\frac{2\pi v_F}{L} \left(L_0-\frac{c}{24}\right)$
is related to the zero mode of the Virasoro stress-tensor, $c$ is the Virasoro central 
charge \cite{CFT-book}, $v_F$ is the Fermi velocity of the edge states and $L$ is the 
circumference of the disk;  the particle number   $Q\equiv N=\sqrt{\nu_H} J_0$ is proportional to 
the zero mode of the $\uu$ current and   $\nu_H$ is the FQH filling factor. 

The Hilbert space $\H_{\mathrm{edge}}$ for the edge-states depends on the number and type of
the residual quasiparticles which might be localized in the bulk when the magnetic field varies slightly
around the value corresponding to the plateau of the Hall conductance. The thermodynamic
parameters, such as the temperature and  the chemical potential 
are related to the \textit{modular parameters} $\t$ and $\z$ on the torus introduced in 
a standard way for the rational CFTs \cite{CFT-book}
\beq \label{modular}
\t=i\pi\frac{T_0}{T}, \quad 
T_0=\frac{\hbar v_F }{\pi k_B L}, \quad \z= i\frac{1 }{2\pi  k_B T} \mu .
\eeq
\subsection{CFT disk partition function  in presence of AB flux}
When magnetic field threading the CB disk or the area of the disk are changed the 
effect on the one-dimensional edge state's 
system\footnote{due to the incompressibility of the FQH droplet, the states in the bulk are localized 
and the only states capable of carrying electric current are living on the edge which is a one-dimensional channel} 
is through the variation of the Aharonov--Bohm (AB) 
flux. As can be seen in Ref.~\cite{NPB-PF_k} introducing AB flux changes the 
boundary conditions of the electron field operator and naturally twists the $\uu$ 
current and the Virasoro stress tensor. The ultimate effect of this twisting on the 
partition function is that it simply shifts the modular parameter as follows
$\z \to \z +\phi \t$, i.e. the partition function in presence of AB flux $\phi$ is
\beq \label{shift}
Z_{\disk}^{\phi}(\t,\z ) = Z_{\disk}(\t,\z +\phi\t).
\eeq
 The Grand potential on the edge \cite{kubo}
\beq\label{omega}\Omega(T,\mu)=-k_B T \ln Z_{\disk}(\t,\z)
\eeq
can be used to compute the particle density in the usual way
\beq \label{density}
\la n \ra_{\b,\mu} = -\frac{k_B T}{L} \frac{\partial}{\partial \mu} \ln Z_{\disk}(\t,\z)=
\frac{1}{L} \la J_0\ra_{\b,\mu}
\eeq
where $\b=(k_B T)^{-1}$ is the inverse temperature and the thermal average is as usual
\beq \label{average}
\la A \ra_{\b,\mu} =  Z_{\disk}^{-1}(\t,\z) \, \mathrm{tr}_{ \H_{\mathrm{edge}}} \  
A\, \e^{2\pi i \t (L_0 -c/24)} e^{2\pi i \z J_0} .
\eeq
\subsection{Coulomb island's  conductance}
In order to obtain the charge stiffness of the CB island, we need to differentiate the particle density
which, according to Eqs.~(\ref{Z}) and (\ref{average}), is related to the 
thermodynamic averages of the zero mode of the $\uu$ current
\beq \label{stiff}
\left\la \frac{\partial n}{\partial \mu} \right\ra_{\b,\mu} = 
\frac{1 }{L k_B T} \left(\la J_0^2\ra_{\b,\mu} -(\la J_0\ra_{\b,\mu})^2 \right) .
\eeq
On the other hand, the Grand potential on the edge $\Omega(T,\mu)$
depends on the AB flux $\phi$ threading the edge because of 
Eqs.~(\ref{modular}) and (\ref{shift}) and the second derivative with respect to $\phi$ is
\beq \label{flux-derivative}
\frac{\partial^2 \Omega}{\partial\phi^2}=-\frac{(hv_F/L)^2}{k_B T}  
\left(\la J_0^2\ra_{\b,\mu} -(\la J_0\ra_{\b,\mu})^2 \right).
\eeq
Comparing Eq.~(\ref{stiff}) with Eq.~(\ref{flux-derivative}) we 
conclude \cite{thermal} that  
\textit{edge conductance is exactly proportional to the magnetic susceptibility} 
$\kappa(T,\phi)=- (e/h)^2 \partial^2 \Omega(T,\phi)/\partial \phi^2$, i.e. 
\beq \label{G_is} 
G_{\mathrm{is}}(T,\phi) =\frac{\s_{\mathrm{is}}(0)}{L}=- \frac{ L }{2v_F} 
\left(\frac{e}{h} \right)^2 \frac{\partial^2 \Omega(T,\phi)}{\partial \phi^2}   
\eeq
This beautiful result, which relates a non-equilibrium quantity, such as the 
CB islands' conductance  $G_{\mathrm{is}}$, to an equilibrium one expressed 
as a derivative 
of the Grand potential $\Omega$, is valid within the Kubo linear response regime, 
characterized by the conditions $G_{L,R}\ll e^2/h$, which is used in the 
derivation \cite{DiVentra} of the Einstein's relation.

\subsection{Disk partition functions for FQH droplets}
To compute the partition function for the edge of a disk FQH sample
we need some knowledge of the structure of the underlying CFT.
The rational CFT for a FQH state always contains a $\uu$ current algebra 
which is completely determined by the filling factor $\nu_H=n_H/d_H$.
This current algebra always contributes a $c=1$ stress-tensor to the 
Virasoro algebra due to the Sugawara contribution  \cite{CFT-book}.
There is in general, a neutral Virasoro generator $T^{(0)}(z)$ as well,
defined by $T(z)-T^{(c)}(z)=T^{(0)}(z)$  whose central charge must be positive.
 
The electron field operator naturally decomposes into a charged $\uu$ part 
and a neutral component which must be a primary field of the neutral Virasoro
algebra. 
From the electron CFT dimension  $\D_{\el}= \frac{d_H}{2n_H}+\D^{(0)}$ 
we see that its statistical angle
$\theta/\pi = 2 \D_\el=2\D^{(0)} + \frac{d_H}{n_H} $, which must be an odd integer,
imposes certain conditions on the structure of the CFT.
In particular, the electron field operator must have a non-trivial neutral component 
when $n_H>1$, hence the neutral Virasoro algebra must be non-trivial, too. 
This also implies that the charged and neutral parts of the RCFT are not 
completely independent and therefore the partition function will not be simply a 
product of charged and neutral partition functions -- instead there are pairing 
rules for the admissible combinations of charged and neutral characters.
\section{Decomposable subalgebra and  $\Z_{n_H}$ grading}
In this section we will consider in more detail the algebraic structure of the rational
CFT corresponding to a general FQH state on a disk.

We start by noting that the $\uu$ part\footnote{this part can be considered as the result of the fusion of the full electron 
operator with its neutral component $\Psi^{(0)}(z)$} of the electron field operator, constructed
as a $\uu$ vertex exponent \cite{fst}, with a charge parameter determined by the filling factor, 
\beq \label{charged}
\np{\psi_\el(z) \overline{\Psi^{(0)}}(z)} \simeq  \np{\e^{ -i\frac{1}{\sqrt{\nu_H}} \phi^{(c)}(z) }} ,
\eeq
of a chiral boson normalized by 
\beq \label{normal}
\la \phi^{(c)}(z)\phi^{(c)}(w)\ra =-\ln (z-w),
\eeq
certainly commutes with all neutral field operators.
However, the vertex exponent (\ref{charged}) has in general a non-integer statistical 
angle $\theta/\pi=d_H/n_H$ and is not local for $n_H>1$. Therefore it does 
not belong to the chiral (super)algebra $\A$ and cannot be used to decompose 
the latter.

The way out of this locality problem is to consider the $n_H$-th power of the 
vertex exponent (\ref{charged})
\beq \label{n-th}
 \np{\exp\left( -i\frac{n_H}{\sqrt{\nu_H}} \phi^{(c)}(z)\right)} =  
\np{\exp\left( -i\sqrt{n_H d_H} \phi^{(c)}(z)\right)} 
\eeq
which still commutes with all neutral field operators but is local because its 
statistics is $\theta/\pi=n_H d_H$, so that it  does belong to $\A$.
It is worth stressing that the $\uu$ vertex operator (\ref{n-th}) together with 
all neutral generators of  $\A$ generates a \textit{decomposable chiral subalgebra}
$\A_D$  of the original chiral superalgebra $\A$
\beq \label{A_D}
\A_D = \uu_{m} \otimes \A^{(0)} \subset \A .
\eeq
We use the notation $\uu_m$ to denote the rational extension \cite{CFT-book,fst} of the $\uu$ 
current algebra with the pair of vertex exponents 
$:\e^{\pm i \sqrt{m} \phi(z)}:$ with $m=n_H d_H$.

Because the decomposable subalgebra $\A_D$ misses only the powers of the full 
electron operator  $\psi_\el^s$ with $s=0,\ldots, n_H-1$, the original superalgebra 
$\A$ can be naturally represented as the following \textit{direct sum decomposition} 
\beq\label{direct}
\A= \mathop{\oplus}\limits_{s=0}^{n_H-1} \psi_\el^s \A_D .
\eeq
Due to the orthogonality of the different powers of the electron field, following 
from the $\uu$ charge conservation, 
\[
\left \langle \psi_\el^s \A_D, \psi_\el^{s'} \A_D \right\rangle = \left \langle \psi_\el^s, \psi_\el^{s'} \right\rangle
\left \langle \A_D,  \A_D \right\rangle = 0  \quad \mathrm{if} \quad s\neq s' ,
\]
where $\la \ldots, \ldots \ra$ denotes the scalar product,
it appears that the decomposition in Eq.~(\ref{direct}) is in fact a 
$\Z_{n_H}$-graded direct sum decomposition.

The virtue of having a decomposable subalgebra is that it defines the following 
\textit{dual algebra inclusion}
\beq \label{dual}
\A_D \subset \A \subset \A^* \subset \A_D^*,
\eeq
which simplifies the construction of the representation spaces. 
It follows from Eq.~(\ref{dual}) that all  representations of $\A$ are also 
representations of $\A_D$ and at the same time that not all representations of 
$\A_D$ are true representations of $\A$.

Given that the decomposable algebra (\ref{A_D}) is simply a tensor product, its irreducible 
representations (IR) are labeled by pairs of quantum numbers $(l,\L)$, where $l$ is the electric 
charge of the bulk quasiparticles in such units that $Q_\el({\textrm{bulk}}) = l/d_H$, and
$\L$ is the (total) neutral topological charge of the bulk quasiparticles.
Then, it follows from Eq.~(\ref{direct}) that all IRs of $\A$ are direct sums of IRs of $\A_D$,
corresponding to the orbit of the simple current's action, hence
we shall be labeling the irreducible representations of $\A$ by the same pair  $(l,\L)$,
corresponding to the $s=0$ component in Eq.~(\ref{direct}).

As follows from Eq.~(\ref{dual}), not all representations of $\A_D$ are 
true representations of the original superalgebra $\A$. In order to identify the 
\textit{physical} excitations, corresponding to the true representations of $\A$
we will require that they are local with respect to the electron field.
The locality principle implies that those IRs of $\A_D$ which are local with 
respect to the electron are also IRs of $\A$. To formulate more precisely  
the locality requirement let us consider the decomposition of the electron field 
and an arbitrary excitation labeled by $(l,\L)$ into $\uu$ and neutral parts
\beqa
\textrm{electron:}\quad\quad  \psi_\el(z) &=&\np{\e^{-i \frac{d_H}{\sqrt{n_H d_H}}\phi^{(c)}(z)}}  \  \Psi^{(0)}_\omega(z)  
  \nn
\quad  \textrm{excitation:} \quad\ \psi_{l,\L}(z) &=&\np{\e^{i \frac{l}{\sqrt{n_H d_H} } \phi^{(c)}(z)}}  \  \Psi^{(0)}_\L(z) , \nonumber
\eeqa
where the $\uu$ boson is normalized as in Eq.~(\ref{normal}), the electric charge
is related to the $\uu$ label $l$ by $Q_\el(l)=l/d_H$, so that the electric charge 
label of the electron is $l=-d_H$, and $\omega$ denotes the (nontrivial) neutral 
topological charge of the electron.

Now, to identify the \textit{physical} excitations within the extended dual algebra
$\A_D^*$ we require local operator product expansion (OPE) of the excitation
with respect to the electron, i.e.,  we require the power of the coordinate distance $(z-w)$ in the
short-distance OPE to be integer 
\[
\psi_\el(z) \psi_{l,\L}(w)  \mathop{\simeq}_{z\to w }
 (z-w)^{-\frac{l}{n_H} + Q_{\omega}(\L)} \quad
\np{\e^{i \frac{l-d_H}{\sqrt{n_H d_H}}\phi^{(c)}(z)}} \quad
\Psi^{(0)}_{\omega*\L}(w),
\]
where $Q_{\omega}(\L)$ is the (neutral) monodromy charge defined by the following combination of 
conformal dimensions $\D_{\L'}$ of the neutral Virasoro IRs
\beq \label{monodromy}
 Q_{\omega}(\L) \equiv \D_{\omega*\L} -\D_\L -\D_\omega \mod \Z, 
\quad \left(\D_\omega = \D^{(0)} \right).
\eeq
Thus, the locality condition implies that the physical excitations (respectively, 
the true IRs of $\A$) must satisfy the following \textit{$\Z_{n_H}$ pairing rule}
which selects the admissible pairs $(l,\L)$ of charged and neutral quantum numbers 
\beq \label{PR}
n_H Q_{\omega}(\L)  \equiv l \mod n_H.
\eeq
The representation spaces of $\A_D=\uu_{m}\otimes \A^{(0)}$ labeled by the pairs 
$(l,\L)$ which obey the PR (\ref{PR}) (that guarantees these pairs are true representations of the 
original algebra $\A$) are naturally tensor products of the representation 
spaces $\H_l^{(c)}$ for the $\uu$ current algebra  and those, $\H_\L^{(0)}$, for the neutral 
Virasoro algebra, i.e.
\beq \label{H_D}
\H_{l,\L}^{D}=\H_l^{(c)}\otimes   \H_\L^{(0)} ,
\eeq
which explains why we looked for a decomposable subalgebra.

The representation spaces  $\H^{\A}_{l,\L}$ for the original algebra $\A$
can be obtained by the action of $\A$ over the lowest-weight state $|l,\L\ra$.
Because of the decomposition (\ref{direct}) this space has a natural direct sum 
decomposition into representation space $\H_{l,\L}^{D}$ for the decomposable subalgebra
\[
\H^{\A}_{l,\L}= \A |l,\L\ra =
\mathop{\oplus}\limits_{s=0}^{n_H-1} \psi_\el^{s}\A^{D}|l,\L\ra =
\mathop{\oplus}\limits_{s=0}^{n_H-1} \J^{s}\H^{D}_{l,\L} ,
\]
where $\J \simeq \psi^*_\el(0)$ is the simple current \cite{CFT-book} representing the action of the 
electron field operator over the lowest-weight states, i.e.
\[
\J | l,\L\ra = | l+d_H, \omega * \L\ra,
\]
which means that the simple current $\J$ acts on lowest-weight states by fusion - the 
$\uu$ charge is simply shifted by the electric charge of the electron, 
while the neutral Virasoro topological charges are fused with that of the electron.

Taking into account Eq.~(\ref{H_D}) we finally obtain the  representation space for $\A$
\beq \label{H}
\H_{l,\L}^{\A}=\mathop{\oplus}\limits_{s=0}^{n_H-1} \J^{s}\left(\H_l^{(c)}\otimes   \H_\L^{(0)} \right)=
\mathop{\oplus}\limits_{s=0}^{n_H-1} \H_{l+s d_H}^{(c)}\otimes   \H_{\omega^s * \L}^{(0)} .
\eeq
The benefit of this representation of the Hilbert space for a general FQH disk is that its $\uu$ part $\H_{l}^{(c)}$,
which is the edge-states' space of the Luttinger liquid, is completely determined by the filling factor $\nu_H$
and the neutral part $\H_\L^{(0)}$ is what distinguishes between FQH states with the same filling factor but 
different universality classes.
\section{The RCFT partition function for a general FQH disk}
Now that we know the general structure of the Hilbert space for an arbitrary FQH disk state we can obtain the 
corresponding structure of the partition function by plugging Eq.~(\ref{H}) into Eq.~(\ref{Z}). 
Notice however, that  the  $\uu_m$ representation 
spaces $\H_l^{(c)}$ entering Eq.~(\ref{H}) correspond to $m=n_H d_H$ and therefore the 
 electric charge operator $Q$ could be represented in terms of  $\uu_m$ number operator $N=J_0/ \sqrt{m}$, i.e.
\beq \label{Q-N}
Q=\sqrt{\frac{n_H}{d_H}}J_0=  \sqrt{\frac{n_H}{d_H}} \sqrt{n_H d_H} N=n_H N.
\eeq
Therefore, using the properties of the trace
as well as the structure of the Hilbert space (\ref{H}), we obtain the main result--the partition function for a general
FQH disk can be represented as a sum of $n_H$ products of $\uu$ and neutral partition functions 
\beq \label{Z-final}
Z_{l,\L}(\t,\z)=\sum_{s=0}^{n_H-1} K_{l+s d_H}(\t,n_H \z; n_H d_H)  \, \ch_{\omega^s *\L}(\t),
\eeq
where the $\uu$ partition functions  $K_{l+s d_H}(\t,n_H \z; n_H d_H) $ are  expressed as 
Luttinger liquid partition functions for $m=n_Hd_H$ in the notation of  \cite{fst}
\beq \label{K}
 K_{l}(\t,\z; m) = \frac{\mathrm{CZ}(\t,\zeta)}{\eta(\t)} \sum_{n=-\infty}^{\infty} 
q^{\frac{m}{2}\left(n+\frac{l}{m}\right)^2}
\e^{2\pi i \z \left(n+\frac{l}{m}\right)} .
\eeq
The absolute temperature and the Bolzmann factor $e^{-\beta}$ are related to the modular parameter 
$\t$
\beq \label{bolzmann}
q=\e^{-\beta\D\varepsilon}=\e^{2\pi i \t}, \quad \D \varepsilon= \hbar\frac{2\pi v_F}{L},
\eeq
where $\D\varepsilon$ is the non-interacting energy spacing, $v_F$ is the Fermi velocity on the edge and 
L is the circumference of the disk. The Dedekind function $\eta$ and Cappelli--Zemba factors \cite{cz}
entering  Eq.~(\ref{K}) are explicitly given by  
\[
 \eta(\t)=q^{1/24}\prod_{n=1}^\infty (1-q^n), \quad \mathrm{CZ}(\t,\zeta)=\e^{-\pi\nu_H\frac{(\Im \z)^2}{\Im\t}}.
\]
It is worth stressing that the $\uu$ partition functions (\ref{K}) are completely explicit and totally determined by 
the filling factor's numerator $n_H$ and denominator $d_H$. The extra $n_H$ in front of $\z$ in the Luttinger-liquid 
partition function  $K_{l+s d_H}(\t,n_H \z; n_H d_H) $ appears due to the relation (\ref{Q-N}).

The neutral partition functions, which are known mathematically as the characters of the representations 
$\H^{(0)}_{\L}$ of the neutral Virasoro algebra with central charge $c-1$, are defined as usual as 
the trace over the representation space \cite{CFT-book}
\[
 \ch_{\L}(\t) = \tr_{\H^{(0)}_{\L}} {q}^{L_0^{(0)}-\frac{c-1}{24}} .
\]
The neutral topological charge of the electron is denoted by $\omega$ and $\omega*\L$ in Eq.~(\ref{Z-final}) denotes
the fusion of the topological charges of the electron and the bulk quasiparticles. Unlike the charged-part partition 
functions the neutral ones are not completely determined by the filling factor, though their structure is almost fixed by the 
neutral weights $\omega$, $\L$ and their fusion rules , thus representing more subtle 
topological properties of the FQH universality class. Fortunately, for most of the FQH universality classes these functions 
are explicitly known.
\section{Application: Coulomb blockade in the $\Z_3$ Read--Rezayi state}
\begin{figure}[htb]
\centering
\includegraphics[bb= 0 0 594 415,clip,width=10.5cm]{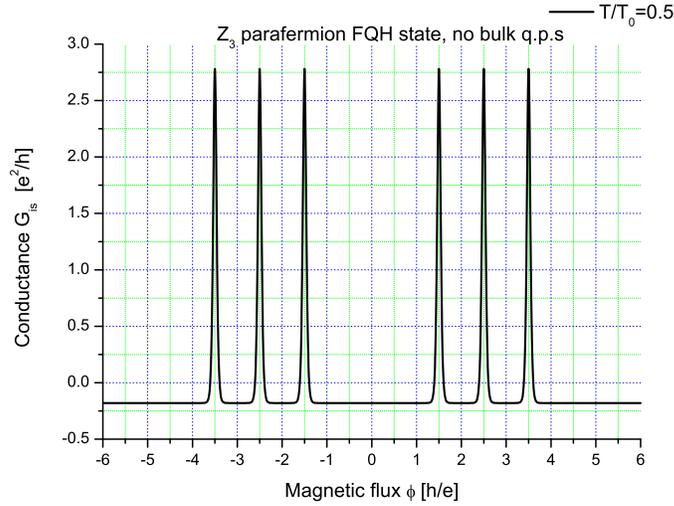}
\caption{Coulomb blockade peaks of the conductance, appearing when AB flux is varied, for the $\Z_3$ 
parafermion FQH island without non-trivial 
quasiparticles in the bulk at temperature $T/T_0=0.5$.  \label{fig:G-PF_3}}
\end{figure}

The structure of the partition function (\ref{Z-final}), in which the $\uu$ part is explicitly separated,
is very convenient for the computation of the CB peaks for a FQH island at finite temperature since 
the variation of the AB flux $\phi$ changes only the $\uu$ partition functions~(\ref{K}) because 
of Eq.~(\ref{shift}). Consider, for example a CB island in which the  FQH state is the $\Z_3$ 
Read--Rezayi (parafermion) state \cite{rr,NPB2001}, characterized by   $n_H=3$, $d_H=5$, i.e. $\nu_H=3/5$.
The decomposable chiral subalgebra is $\widehat{u(1)}_{15}\times{\W_3}$, where $\W_3$ is the 
$\Z_3$ parafermion algebra of Fateev-Zamolodchikov \cite{fat-zam}. The neutral part of the electron operator
has a topological charge $\omega=\psi_1$, \ $\omega^2=\psi_2$  given by the parafermion currents.
As a simple illustration of the entire procedure let us consider the case when there are no 
quasiholes in the bulk,  which corresponds to $l=0$, $\L=0$.  The partition function (\ref{Z-final})
takes the form
\[
Z_{0,0}(\t,\z)=K_0(\t;3\z;15) \mathrm{ch}_{00}(\t) +K_5(\t;3\z;15) \mathrm{ch}_{01}(\t) 
 +K_{-5}(\t;3\z;15) \mathrm{ch}_{02}(\t) 
\]
where the $K$ functions are defined in Eq.~(\ref{K}), the Bolzmann factor $q$ is defined in 
Eq.~(\ref{bolzmann})  and the  neutral partition functions are defined by
\[
\ch_{0,l}(\t) = q^{-\frac{1}{30}} \sum_{n_1, \ n_2\geq 0 }^{(l)}
\frac{q^{\frac{2}{3}\left(n_1^2+n_2n_2+n_2^2 \right)}}{(q)_{n_1}(q)_{n_2}}, 
\quad  (q)_n=\prod_{j=1}^n (1-q^j) .
\]
and the sum $\sum^{(l)}$ is restricted by the condition
 ${n_1+2n_2=l \mod 3}$.
 Introducing AB flux as in Eq.~(\ref{shift}) and plugging the partition function with flux into Eq.~(\ref{G_is})
 we calculate numerically the conductance of the CB island at temperature $T=0.5 T_0$ as the flux is varied, 
see Fig.~\ref{fig:G-PF_3}.
Under the assumption that the neutral and charged modes propagate with the same Fermi velocity 
we see that the CB peaks are clustered in bunches of three, separated by flux period $\D\phi_1 =1$ inside the bunch, 
and separated by a flux period $\D\phi_2 =3$ between the bunches, which is in agreement with the previous results 
at zero temperature \cite{Stern-CB-RR,nayak-doppel-CB,cappelli-viola-zemba}.

Most of the characteristics of the CB peaks, such as the height, the width and the periods, can be derived
asymptotically at very low temperatures \cite{thermal}.
\begin{acknowledgement}
I would like to thank Andrea Cappelli, Guillermo Zemba, Ady Stern, Pasquale Sodano  and Reinhold Egger for many useful discussions
as well as INFN-Firenze and the Galileo Galilei Institute for Theoretical Physics for hospitality and support.
The author has been supported as a Research Fellow by 
the Alexander von Humboldt Foundation as well as by ESF and by the BG-NCSR under Contract DO~02-257.
\end{acknowledgement}


\biblstarthook{}

\providecommand{\href}[2]{#2}\begingroup\raggedright\endgroup


\end{document}